\documentstyle{article}
\title{\bf
Explicit proof of equivalence of two-point functions in the two formalisms
of thermal field theory
\thanks{This project was supported partially by National Natural Science 
Foundation of China and by Grant No.LWTZ-1298 of the Chinese Academy of 
Sciences.}}
\author{
{\bf 
Bang-Rong Zhou} \\
\normalsize Department of Physics, Graduate School, Academia Sinica,
Beijing 100039, {\bf China} \\
\normalsize and  \\
\normalsize CCAST ( World Laboratory ) P.O.Box 8730, Beijing 100080,
{\bf China}\\
}
\date {}
\begin{document}
\newcommand{\dfrac}[2]{\frac{\displaystyle #1}{\displaystyle #2}}
\hoffset = -1 truecm
\voffset = -2 truecm
\baselineskip = 12pt
\maketitle
\begin{abstract}
We give an explicit proof of equivalence of the two-point function to
one-loop order in the two formalisms of thermal $\lambda \varphi^3$ theory
based on the expressions in the real-time formalism. It is indicated that
the key-point of completing the proof is to separate carefully the imaginary
part of the zero-temperature loop integral from relevant expressions and
this fact will certainly be very useful for examination of the equivalence
problem of the two formalisms of thermal field theory in other theories,
including the one of the propagators for scalar bound states in a NJL  model.
\end {abstract}
PACS numbers: 11.10.Wx, 11.10.-z, 03.70.+k\\
Key words: two-point function, thermal field theory, equivalence of the 
imaginary-time and real-time formalism, imaginary part of zero-temperature loop
%%-main body of paper-%%
\section {Introduction}
\indent  Finite temperature field theory has attracted much research interests
owing to its application to phase transition of early universe and nuclear
matter [1-5]. It has two formalisms: the imaginary-time and the real-time
formalism [4].  However, the complete equivalence between the two formalisms
has been a subtle problem, though one generally believes that the two formalisms
should give identical results [6]. In a recent research on the Nambu-Goldstone
mechanism of electroweak symmetry breaking at finite temperature [7], we
calculate the propagators for scalar bound states in one-generation fermion
condensate model [a Nambu-Jona-Lasinio (NJL)-type model [8]] and find that
their denominators show different imaginary-parts in the two formalisms.
By conventional inference, origin of the difference could be that either we
have calculated different Green functions in the two formalisms [9], or we have
missed some important technical details in the calculations. To clarify matter,
we will review some similar calculations. Because the propagators for scalar
bound states in a NJL model correspond to four-point amputated functions and the
calculations of them could be effectively reduced to the ones of usual two-point
functions, we will look back on the discussions of some two-point functions.
Among them, the most typical and simple one is the  two-point function in
$\lambda\varphi^3$ theory.  It is accepted that  two-point functions in the
imaginary-time and the real-time formalism is equivalent.  This has been
discussed by a formal analysis [4] and some calculations of the imaginary parts
of the Green functions [10-11]. However, as far as the whole two-point  Green
function is concerned, no explicit calculation of this equivalence based on the
expressions in the real-time formalism has been given. In fact, for the sake of
comparison with four-point amputated functions in a NJL model we need   the
calculations of the whole two-point function and in particular, expect to
express the results in the real-time formalism so as to show correspondence
between a causal Green function and a physical propagator.
Following this idea, in this paper we will give an explicit proof of
equivalence of  the whole two-point function  to one-loop order in
$\lambda \varphi^3$ theory in the two formalisms by means of the expressions in
the real-time formalism and from it find the key-point of completing the proof.
This  will certainly be  very useful for reexamination of the results in a NJL 
model.\\
\indent The paper is arranged as follows.  In Sect.2 we will calculate the
two-point function to one-loop order in the imaginary-time formalism, then
analytically continue the function defined at the Matsubara frequency to the
physical causal propagator defined  at real energy.  In Sect.3  the
calculations of the two-point function in the real-time formalism will be
conducted. After diagonalization of the two-point function matrix by means of
a thermal transformation matrix, we will obtain the physical causal propagator
in the real-time formalism. In Sect.4 we will explicitly prove that the causal
propagators obtained in the two formalisms are identical and indicate that
the key-point of completing the proof is to separate carefully  the imaginary
part of the zero-temperature loop integral. Finally, in Sect.5  we come to our
conclusions.
\section{Two-point function in the imaginary-time formalism}
\indent In the imaginary-time formalism, Lagrangian can be expressed by
$${\cal L}=
\frac{1}{2}\partial_{\mu}\varphi\partial^{\mu}\varphi-
\frac{1}{2}m^2\varphi^2-\frac{\lambda}{3!}\varphi^3.
\eqno(1)$$
The vertex rule will be $-\lambda$ and the propagator for a free particle will
be $\Delta_F(i\omega_n,\stackrel{\rightharpoonup}{l})=1/(\omega^2_n+\omega^2_l)$,
where $\omega_n=2\pi n/\beta (n=0,\pm1,\pm2, ...)$ is the Matsubara frequency ,
$\beta=1/T$ is the reciprocal of temperature and
$\omega^2_l={\stackrel{\rightharpoonup}{l}}^2+m^2$ with the three momentum
$\stackrel{\rightharpoonup}{l}$ [4,5,12]. The two-point function
$G_F^I(-i\omega_m, \stackrel{\rightharpoonup}{p})$ to one-loop order submits
to the equation
$$ G_F^I(-i\omega_m, \stackrel{\rightharpoonup}{p})=
\Delta_F(-i\omega_m, \stackrel{\rightharpoonup}{p})+
\Delta_F(-i\omega_m, \stackrel{\rightharpoonup}{p})
\Pi^I(-i\omega_m, \stackrel{\rightharpoonup}{p})
G_F^I(-i\omega_m, \stackrel{\rightharpoonup}{p}), \ \ \
\omega_m=\frac{2\pi m}{\beta},
\eqno(2)$$
which has the solution
$$G_F^I(-i\omega_m, \stackrel{\rightharpoonup}{p})=
\frac{1}{\omega_m^2+\omega_p^2-\Pi^I(-i\omega_m, \stackrel{\rightharpoonup}{p})},
 \ \ \ \omega_p^2=
{\stackrel{\rightharpoonup}{p}}^2+m^2 ,\eqno(3)$$
where $\Pi^I(-i\omega_m, \stackrel{\rightharpoonup}{p})$ is the contribution
from one-loop diagram and may be expressed by
$$\Pi^I(-i\omega_m, \stackrel{\rightharpoonup}{p})
=\lambda^2\int\frac{d^3l}{(2\pi)^3}A(-i\omega_m, \stackrel{\rightharpoonup}{p},
\stackrel{\rightharpoonup}{l})
\eqno(4)$$
with
$$ A(-i\omega_m, \stackrel{\rightharpoonup}{p},\stackrel{\rightharpoonup}{l})
=T\sum_n\frac{1}{\omega^2_n+\omega^2_l}
\frac{1}{(\omega_m+\omega_n)^2+\omega^2_{l+p}}, \ \ \ 
\omega^2_{l+p}=
(\stackrel{\rightharpoonup}{l}+\stackrel{\rightharpoonup}{p})^2+m^2.
\eqno(5)$$
The sum of the Matsubara frequency in Eq. (5) can be done by standard
procedure [12].  We define the Fourier transformation
$$\frac{1}{\omega^2_m+\omega^2_k}=\int_0^{\beta}\tilde{\Delta}_F(\tau,\omega_k)
e^{i\omega_m\tau}\eqno(6)$$
and the inverse formula
$$\tilde{\Delta}_F(\tau,\omega_k)=T\sum_ne^{-i\omega_n\tau}
\frac{1}{\omega^2_n+\omega^2_k} 
\eqno(7)$$
which obeys the periodicity condition
$$\tilde{\Delta}_F(\tau,\omega_k)= \tilde{\Delta}_F(\tau - \beta,\omega_k)
\eqno(8)$$
and can be calculated by the formula
$$\tilde{\Delta}_F(\tau,\omega_k)=T\sum_n g(k_0=i\omega_n)
=T\int_{C_1\cup C_2}\frac{dk_0}{2\pi i}
g(k_0)\frac{\beta}{2}\coth\frac{\beta k_0}{2},$$ 
$$
g(k_0)=e^{-k_0\tau}\frac{1}{\omega^2_k-k^2_0}, \ \ \ {\rm for} \ \ \ \tau >0,
\eqno(9)$$
where$C_1$ and $C_2$ represent the integral paths $\eta -i\infty \to
\eta+i\infty$ and $-\eta+i\infty \to -\eta-i\infty (\eta=0_+)$ respectively in
the complex $k_0$ plane.  The result is
$$\tilde{\Delta}_F(\tau,\omega_k)=\frac{1}{2\omega_k}
\{n(\omega_k)e^{\omega_k \tau}+[1+n(\omega_k)] e^{-\omega_k \tau}\},
\eqno(10)$$
$$n(\omega_k)=1/(e^{\beta\omega_k}-1).
\eqno(11)$$
By means of Eqs. (6),(10) and the completeness formula
$$T\sum_ne^{i\omega_n(\tau-{\tau}')}=\delta(\tau-{\tau}'),
\eqno(12)$$
we may obtain the frequency sum (5) expressed by
$$
\begin{array}{rcl}
 A(-i\omega_m, \stackrel{\rightharpoonup}{p},\stackrel{\rightharpoonup}{l})&=&
   \int_0^{\beta}d\tau\tilde{\Delta}_F(\tau,\omega_l) 
\tilde{\Delta}_F(\tau,\omega_{l+p})e^{-i\omega_m\tau} \vspace{0.5cm}\\
&=&\dfrac{1}{4\omega_l \omega_{l+p} }\left\{
\dfrac{1+n(\omega_l)+n(\omega_{l+p})}{i\omega_m+\omega_l+\omega_{l+p}}+
\dfrac{ n(\omega_{l+p})- n(\omega_l)}{i\omega_m+\omega_l-\omega_{l+p}} \right. 
\vspace{0.5cm}\\
& & \ \ \ \ \ \ \ 
\left.+\dfrac{ n(\omega_l)-n(\omega_{l+p})}{ i\omega_m-\omega_l+\omega_{l+p}}-
     \dfrac{1+n(\omega_l)+n(\omega_{l+p})}{ i\omega_m-\omega_l-\omega_{l+p}}
\right\}.   \vspace{0.5cm}
\end{array}\eqno(13)$$
This completes the calculation of 
$ G_F^I(-i\omega_m, \stackrel{\rightharpoonup}{p})$ in Eq. (3).\\
\indent To obtain physical causal Green function, we must make the analytic 
continuation of the energy from the discrete imaginary values to the real axis
in the following way [12]
$$-i\omega_m \rightarrow  p^0+ i \varepsilon p^0, \ \  \varepsilon =0_+ .
\eqno(14) $$
Taking an additional factor $-i$  into account, we may express the physical
causal two-point function by
$$G^I_F(p)=\left.-iG^I_F(-i\omega_m, \stackrel{\rightharpoonup}{p})
\right |_{-i\omega_m \to p^0+ i \varepsilon p^0}
=i/[p^2-m^2+\Pi^I(p)+i\varepsilon],
\eqno(15)$$
noting that under the continuation (14), 
$\omega_m^2+\omega^2_p\to -(p^2-m^2+i\varepsilon)$, and
$$\Pi^I(p)=\lambda^2\int\frac{d^3l}{(2\pi)^3}
A_F(p,\stackrel{\rightharpoonup}{l}),
\eqno(16) $$
$$\begin{array}{rcl}
A_F(p, \stackrel{\rightharpoonup}{l})&=& A(-i\omega_m, 
\stackrel{\rightharpoonup}{p},\stackrel{\rightharpoonup}{l})|_{-i\omega_m \to 
p^0+ i \varepsilon \eta(p^0)} \vspace{0.5cm}\\
&=&\dfrac{1}{4\omega_l \omega_{l+p} }\left[
\dfrac{1+n(\omega_l)+n(\omega_{l+p})}{-p^0+\omega_l+\omega_{l+p}
-i\varepsilon\eta(p^0)}+\dfrac{ n(\omega_{l+p})- 
n(\omega_l)}{-p^0+\omega_l-\omega_{l+p}-i\varepsilon\eta(p^0)}\right. 
\vspace{0.5cm}\\
&&\left.+\dfrac{ n(\omega_l)-n(\omega_{l+p})}{ -p^0-\omega_l+\omega_{l+p}- 
i\varepsilon\eta(p^0)}-
\dfrac{1+n(\omega_l)+n(\omega_{l+p})}{ -p^0-\omega_l-\omega_{l+p}- 
i\varepsilon\eta(p^0)}
\right], \ \ \eta(p^0)=\dfrac{p^0}{|p^0|} .
\end{array}\eqno(17) $$
For making a comparison with the following results obtained in the real-time
formalism, we change $ A_F(p, \stackrel{\rightharpoonup}{l})$ into  an integral
representation. By the formula
$$\frac{1}{X+i\varepsilon}=\frac{X}{X^2+\varepsilon^2}- i \pi\delta(X)
\eqno(18)$$
and the definition
$$n(l^0)=\theta(l^0)/(e^{\beta l^0}-1)+  \theta(-l^0)/(e^{-\beta l^0}-1)
        =1/(e^{\beta |l^0|}-1)
        =\sinh^2\Theta(l^0),
\eqno(19)$$
we may write
$$\begin{array}{rcl}
A_F(p, \stackrel{\rightharpoonup}{l})&=&
\int\dfrac{dl^0}{2\pi}\dfrac{-i}{({l^0}^2
-\omega_l^2+i\varepsilon)[(l^0+p^0)^2-\omega^2_{l+p}+i\varepsilon]}  
\vspace{0.5cm}\\
&&-\int dl^0\left\{ \sinh^2\Theta(l^0)
{\rm P}\dfrac{\delta({l^0}^2-\omega^2_l)}{ (l^0+p^0)^2-\omega^2_{l+p}}+ 
\sinh^2\Theta(l^0+p^0){\rm P}\dfrac{\delta[(l^0+p^0)^2-\omega^2_{l+p}]}
{ {l^0}^2-\omega^2_l}
 \right\} \vspace{0.5cm}\\
&&+ i \eta(p^0)\pi\int dl^0\delta({l^0}^2-\omega^2_l) 
\delta[(l^0+p^0)^2-\omega^2_{l+p}] \end{array}$$
$$ \cdot [\sinh^2\Theta(l^0)\eta(l^0+p^0)+\sinh^2\Theta(l^0+p^0)
\eta(-l^0)] ,\eqno(20)$$
where $"{\rm P}"$ means the principal value of the integral.
Substituting Eq.(20) into Eq.(16), we obtain
$$\Pi^I(p)=\tilde{K}(p)-\tilde{H}(p)+i\tilde{S}^I(p),
\eqno(21)$$
where
$$\tilde{K}(p)=\lambda^2\int\frac{d^4l}{(2\pi)^4}
\frac{-i}{(l^2-m^2+i\varepsilon)[(l+p)^2-m^2+i\varepsilon]}=
\frac{\lambda^2}{16\pi^2}\int_0^1dx\left(\ln\frac{\Lambda^2+M^2}{M^2}
-\frac{\Lambda^2}{\Lambda^2+M^2}\right)
\eqno(22)$$
with $M^2=m^2-p^2x(1-x)$ and the Euclidean four-momentum cutoff $\Lambda$,
is the zero-temperature loop integral,
$$\tilde{H}(p)=2\pi\lambda^2\int\frac{d^4l}{(2\pi)^4}\left\{\frac{(l+p)^2-m^2}
{[(l+p)^2-m^2]^2+\varepsilon^2}+(p\to -p)\right\}
\delta(l^2-m^2)\sinh^2\Theta(l^0)
\eqno(23)$$
and
$$\tilde{S}^I(p)=\eta(p^0)2\pi^2\lambda^2\int\frac{d^4l}{(2\pi)^4} 
\delta(l^2-m^2)\delta[(l+p)^2-m^2][\sinh^2\Theta(l^0)\eta(l^0+p^0)
+\sinh^2\Theta(l^0+p^0)\eta(-l^0)].
\eqno(24)$$
Eq. (15) together with Eqs. (21)-(24) give physical causal two-point function
in the imaginary-time formalism.
\section{Two-point function in the real-time formalism}
\indent In the real-time formalism, Lagrangian can be expressed by
$${\cal L}=
\frac{1}{2}\sum_{a=1,2}\partial_{\mu}\varphi^{(a)}\partial^{\mu}\varphi^{(a)}
-\frac{1}{2}m^2\sum_{a=1,2}\varphi^{(a)} 
\varphi^{(a)}-\frac{\lambda}{3!}\sum_{a=1,2}(-1)^{a+1}[\varphi^{(a)}]^3 ,
\eqno(25)$$
where   $a=1$ and $a=2$ respectively represent physical and ghost field. The
vertex rule will be $-i\lambda(-1)^{a+1}$ and the propagator for a free
particle will become a matrix  $D^{ab}(p) (a,b=1,2)$ [4] with the elements
expressed by
$$D^{11}(p)=\frac{i}{p^2-m^2+i\varepsilon}+n(p^0)2\pi \delta(p^2-m^2)
           =[D^{22}(p)]^*, $$ $$
D^{12}(p)=D^{21}(p)=e^{\beta|p^0|} n(p^0)2\pi \delta(p^2-m^2).
\eqno(26)$$
The two-point function ${G^R}^{ab}(p) (a,b=1,2)$ obeys the equation
$$G^{Rab}(p)=D^{ab}(p)+D^{ac}(p)i\Pi^{cd}(p)G^{Rdb}(p),
\eqno(27)$$
where $i\Pi^{cd}(p)$ is the contribution from the one-loop diagram  with the
vertex denotations $c$ and $d$.  The matrix form of Eq.(27) is
$$G^R(p)=D(p)+D(p)i\Pi(p)G^R(p) .
\eqno(28)$$
The propagator matrix $D(p)$ may be diagonalized by a thermal transformation
matrix $M$ [4], i.e.
$$D(p)=M \left(\matrix{\Delta_F(p) & 0            \cr
0 & \Delta_F^*(p)\cr}\right) M,\ \ 
1 \Delta_F(p)=\frac{i}{p^2-m^2+i\varepsilon},$$ $$
        M=\left(\matrix{\cosh\Theta(p^0) & \sinh\Theta(p^0)\cr
                        \sinh\Theta(p^0) & \cosh\Theta(p^0)\cr}
           \right), \sinh^2\Theta(p^0)=n(p^0).
\eqno(29)$$
The same matrix $M$ may diagonalize the two-point function matrix $G^R(p)$
and give the physical causal two-point function $G_F^R(p)$ [4].  Thus we
assume  that
$$G^R(p)=M \left(\matrix{ G_F^R(p) & 0            \cr
                             0     & {G_F^R}^*(p)\cr}\right) M
\eqno(30)$$
and
$$i\Pi(p)=M^{-1}\left(\matrix{ i\Pi^R(p) & 0            \cr
                             0     & -i{\Pi^R}^*(p)\cr}\right) M^{-1} ,
\eqno(31)$$
then may obtain from Eq.(28) that
$$
\begin{array}{rcl}
\left(\matrix{ G_F^R(p) & 0            \cr
                 0        & {G_F^R}^*(p)  \cr}\right)&=&
  \left(\matrix{\Delta_F(p) & 0            \cr
                 0          & \Delta_F^*(p)\cr}\right) \vspace{0.5cm}\\
  &&+\left(\matrix{\Delta_F(p) & 0            \cr
                 0          & \Delta_F^*(p)\cr}\right)
  \left(\matrix{ i\Pi^R(p)  & 0            \cr
                 0          & -i{\Pi^R}^*(p)\cr}\right)
  \left(\matrix{ G_F^R(p)   & 0            \cr
                 0          & {G_F^R}^*(p)\cr}\right)
\end{array}\vspace{0.5cm}
\eqno(32)$$
which has the solution
$$G^R_F(p)=\frac{1}{\Delta^{-1}_F(p)-i\Pi^R(p)}=
\frac{i}{p^2-m^2+i\varepsilon+\Pi^R(p)}.
\eqno(33)$$
It is indicated that in deriving Eq.(33) we have used the diagonalization
assumption (31) of $i\Pi(p)$ and this means that
$$ \left(\matrix{\Pi^R(p)  & 0            \cr
                  0        & -{\Pi^R}^*(p)\cr}\right)=M\Pi(p)M=
   M\left(\matrix{\Pi^{11}  & \Pi^{12}     \cr
                  \Pi^{21}  & \Pi^{22}     \cr}\right)M .
\eqno(34)$$
By means of the explicit expression of $M$ in Eq. (29), Eq. (34) will be
reduced to the following constraints:
$$\Pi^{11}+\Pi^{22}+ e^{\beta|p^0|/2}\Pi^{12}+ e^{-\beta|p^0|/2}\Pi^{21}=0, $$
$$
\Pi^{11}+\Pi^{22}+ e^{-\beta|p^0|/2}\Pi^{12}+ e^{\beta|p^0|/2}\Pi^{21}=0
\eqno(35)$$
and
$$\Pi^R=\cosh^2\Theta\Pi^{11}+\sinh^2\Theta\Pi^{22}+\cosh\Theta\sinh\Theta
(\Pi^{12}+\Pi^{21}),$$ $$
-{\Pi^R}^*=\sinh^2\Theta\Pi^{11}+\cosh^2\Theta\Pi^{22}+\sinh\Theta\cosh\Theta 
(\Pi^{12}+\Pi^{21}).
\eqno(36)$$
Since the two equations in Eq.(35) exchange each other under
$|p^0| \leftrightarrow -|p^0|$, we may remove the absolute value symbol of
$p^0$, i.e.  write
$$\Pi^{11}+\Pi^{22}+ e^{\beta p^0/2}\Pi^{12}+ e^{-\beta p^0/2}\Pi^{21}=0, $$
$$
\Pi^{11}+\Pi^{22}+ e^{-\beta p^0/2}\Pi^{12}+ e^{\beta p^0/2}\Pi^{21}=0.
\eqno(37)$$
On the other hand, Eq.(36) will lead to that
$$\Pi^{11}=-{(\Pi^{22})}^*, \ \ \Pi^{12}+\Pi^{21}=-(\Pi^{12}+\Pi^{21})^*.
\eqno(38)$$
Determination of $G^R_F(p)$ depends on calculation of $\Pi^R(p)$ and in the
same time we must check validity of the relations (37) and (38).  In the
one-loop  approximation,
$$i\Pi^{cd}(p) =
(-1)^{c+d+1}\lambda^2\int\frac{d^4l}{(2\pi)^4}D^{cd}(l+p)D^{dc}(l)
\eqno(39)$$
with the following explicit expressions
$$\Pi^{11}(p)=\tilde{K}(p)-\tilde{H}(p)+i \tilde{S}(p),$$ 
$$\Pi^{22}(p)=-{\tilde{K}}^*(p)+\tilde{H}(p)+i \tilde{S}(p),$$
$$\Pi^{12}(p)= \Pi^{21}(p)=-i \tilde{R}(p),
\eqno(40)$$
where $\tilde{K}(p)$ and $\tilde{H}(p)$ are given by Eqs.(22) and (23) , and
$$\tilde{S}(p)= 2\pi^2\lambda^2\int\frac{d^4l}{(2\pi)^4} \delta(l^2-m^2) 
\delta[(l+p)^2-m^2]
$$
$$\cdot [\sinh^2\Theta(l^0)\cosh^2\Theta(l^0+p^0)+
\sinh^2\Theta(l^0+p^0)\cosh^2\Theta(l^0)],
\eqno(41)$$
$$\tilde{R}(p)= \pi^2\lambda^2\int\frac{d^4l}{(2\pi)^4} \delta(l^2-m^2) 
\delta[(l+p)^2-m^2]\sinh 2\Theta(l^0)\sinh 2\Theta(l^0+p^0).
\eqno(42)$$
Obviously, $\Pi^{cd}(p)$ given by Eq.(40) satisfy Eq.(38) and since 
$\Pi^{12}=\Pi^{21}$, Eq. (37) is reduced to
$$\Pi^{11}+\Pi^{22}+\cosh(\beta p^0/2)\Pi^{12}=0$$
and furthermore to
$$\tilde{S}(p)= \cosh(\beta p^0/2)\tilde{R}(p)-{\rm Im}\tilde{K}(p),
\eqno(43)$$
where ${\rm Im}\tilde{K}(p)$ represents the imaginary part of $\tilde{K}(p)$.
To verify validity of Eq.(43), we will use the explicit expressions (41) and
(42). In fact, by means of the definition (19), we can rewrite $\tilde{R}(p)$
and $\tilde{S}(p)$ by
$$\tilde{R}(p)=
\pi^2\lambda^2\int\frac{d^4l}{(2\pi)^4} \delta({l^0}^2-\omega_l^2) 
\delta[(l^0+p^0)^2-\omega_{l+p}^2]
\frac{\eta(l^0+p^0)\eta(l^0)}{\sinh[\beta(l^0+p^0)/2] \sinh(\beta l^0/2)}
\eqno(44)$$
and
$$\tilde{S}(p)= \cosh(\beta p^0/2)\tilde{R}(p)-\tilde{\Delta}(p),
\eqno(45)$$
where
$$\begin{array}{rcl}
\tilde{\Delta}(p)&=&2\pi^2\lambda^2\int\dfrac{d^4l}{(2\pi)^4} 
\delta({l^0}^2-\omega_l^2) \delta[(l^0+p^0)^2-\omega_{l+p}^2]
[\theta(-l^0)\theta(l^0+p^0)+\theta(l^0)\theta(-l^0-p^0)]\\
&=&2\pi^2\lambda^2\int\dfrac{d^3l}{(2\pi)^4}\dfrac{1}{4\omega_l\omega_{l+p}}
 [\delta(p^0+\omega_l+\omega_{l+p})+\delta(p^0-\omega_l-\omega_{l+p})].
\end{array}\eqno(46)$$
On the other hand, we can calculate $\tilde{K}(p)$ from Eq.(22) by the residue
theorem and obtain
$$\tilde{K}(p)=
2\pi\lambda^2\int\frac{d^3l}{(2\pi)^4}\frac{1}{4\omega_l\omega_{l+p}}
\left[\frac{1}{ p^0+\omega_l+\omega_{l+p}-i\varepsilon}-
\frac{1}{ p^0-\omega_l-\omega_{l+p}+i\varepsilon}\right].
\eqno(47)$$
Thus the imaginary part of $\tilde{K}(p)$ may be derived and the result is
$${\rm Im}\tilde{K}(p)=
2\pi^2\lambda^2\int\frac{d^3l}{(2\pi)^4}\frac{1}{4\omega_l\omega_{l+p}}
   [\delta(p^0+\omega_l+\omega_{l+p})+\delta(p^0-\omega_l-\omega_{l+p})]
   =\tilde{\Delta}(p).
\eqno(48)$$
This verifies that Eq. (43) thus Eq.(37) is satified indeed. 
It should be emphasized that for verification of Eq.(43), separation of 
$-{\rm Im}\tilde{K}(p)$ from $\tilde{S}(p)$ is essential. Substituting Eq. (40)
 into Eq. (36) and using Eqs. (45), (48) and (19), we obtain
$$\Pi^R(p)= \tilde{K}(p)- \tilde{H}(p)+ i[\sinh(\beta |p^0|/2)\tilde{R}(p)
-{\rm Im}\tilde{K}(p)].
\eqno(49)$$
Eq. (33) together with Eq. (49) give the causal two-point function to one-loop
order in the real-time formalism.
\section{Equivalence of two-point functions in the two formalisms}
\indent If the causal two-point functions calculated in the two formalisms are 
equivalent, then we must have $G^{I}_F(p)=G^R_F(p)$, and from Eqs.(15) and (33),
this means that $\Pi^I(p)=\Pi^R(p)$, or explicitly, by Eqs.(21) and (49),
$$\tilde{K}(p)-\tilde{H}(p)+i\tilde{S}^I(p)= 
\tilde{K}(p)- \tilde{H}(p)+ i[\sinh(\beta |p^0|/2)\tilde{R}(p)-
{\rm Im}\tilde{K}(p)].
\eqno(50)$$
Since the real parts [$Re\tilde{K}(p)- \tilde{H}(p)$] in the two sides are
the same, the problem is reduced to proving that
$$\tilde{S}^I(p)+ {\rm Im}\tilde{K}(p)=\sinh(\beta |p^0|/2)\tilde{R}(p),
\eqno(51)$$
i.e. $\Pi^I(p)$ and $\Pi^R(p)$ must have identical imaginary parts. To verify
Eq.(51), we use the expression (24) of $\tilde{S}^I(p)$, the definition (19)
and Eq.(44) and get
$$\tilde{S}^I(p)=\eta(p^0) \sinh(\beta p^0/2)\tilde{R}(p)-\tilde{\Delta}_1(p),
\eqno(52)$$
where
$$\tilde{\Delta}_1(p)=\eta(p^0) \pi^2\lambda^2\int\frac{d^4l}{(2\pi)^4} 
\delta(l^2-m^2) \delta[(l+p)^2-m^2][\eta(l^0+p^0)-\eta(l^0)].
\eqno(53)$$
Owing to the factor $\eta(l^0+p^0)-\eta(l^0)$, in the integrand of Eq. (53) only
the terms containing $\delta(l^0\mp \omega_l)\delta(l^0+p^0\pm \omega_{l+p})$
have non-zero contributions,  we thus obtain
$$\begin{array}{rcl}
\tilde{\Delta}_1(p)&=&\eta(p^0)2 \pi^2\lambda^2\int\dfrac{d^3l}{(2\pi)^4}
   \dfrac{1}{4\omega_l\omega_{l+p}}
   [-\delta(p^0+\omega_l+\omega_{l+p})+\delta(p^0-\omega_l-\omega_{l+p})] \\
&=&2 \pi^2\lambda^2\int\dfrac{d^3l}{(2\pi)^4}
   \dfrac{1}{4\omega_l\omega_{l+p}}
   [\delta(p^0+\omega_l+\omega_{l+p})+\delta(p^0-\omega_l-\omega_{l+p})] \\
&=&{\rm Im}\tilde{K}(p). 
\end{array} \eqno(54)$$
Substituting Eq. (54) into Eq. (52) will mean that Eq. (51) is proven.
Here we also indicate that separation of $-{\rm Im}\tilde{K}(p)$ from 
$\tilde{S}^I(p)$ is essential for verification of Eq. (51).
This result shows the two-point function $G^I_F(p)$ in Eq.(15) and
$G^R_F(p)$ in Eq. (33) are actually identical , hence we may omit the
superscripts $"I"$ and $"R"$ and simply express the physical causal two-point
function by
$$\begin{array}{rcl}
G_F(p)&=&G^I_F(p)=G^R_F(p) \\
&=& i /\{p^2-m^2+ i \varepsilon + {\rm Re}\tilde{K}(p)-\tilde{H}(p)+ 
                             i[\tilde{S}^I(p)+ {\rm Im}\tilde{K}(p) ]\} \\
&=& i /\{p^2-m^2+ i \varepsilon + {\rm Re}\tilde{K}(p)-\tilde{H}(p)+ 
                             i\sinh(\beta |p^0|/2)\tilde{R}(p)\}. \\
\end{array}\eqno(55)$$
From the above demonstration we see that a key-point of completing the
equivalence proof is carefully to consider and separate the imaginary part
${\rm Im}\tilde{K}(p)$ (with a minus sign) of the zero-temperature loop
integral from the relevant expressions e.g. $\tilde{S}(p)$ and $\tilde{S}^I(p)$.
Without doing so, we can not explain the diagonalization of the matrix $\Pi(p)$
in the real-time formalism, and in particular, can not prove the identity of the
imaginary parts of the denominator of $G_F(p)$ in the two formalisms. \\
\indent By means of the causal two-point function $G_F(p)$, we can also obtain
the retarded and advanced two-point functions $G_r(p)$ and $G_a(p)$.  In fact,
if starting from the imaginary-time formalism, then we may define [12]
$$G_F^I(p)=-iG(- i\omega_m \to p^0+ i\varepsilon 
p^0,\stackrel{\rightharpoonup}{p}),$$
$$ G_r^I(p)=-iG(-i\omega_m \to p^0+ 
i\varepsilon ,\stackrel{\rightharpoonup}{p}),$$
$$ G_a^I(p)=-iG(- i\omega_m \to p^0- i\varepsilon ,
\stackrel{\rightharpoonup}{p}).
\eqno(56)$$
From Eq. (56) we derive that
$$ G_F^I(p)= \theta(p^0)G_r^I(p)+\theta(-p^0)G_a^I(p),
\eqno(57)$$
$$  [G_r^I(p)]^*=- G_a^I(p),
\eqno(58)$$
and then
$$ G_r^I(p)= \theta(p^0)G_F^I(p)-\theta(-p^0)[G_F^I(p)]^*.
\eqno(59)$$
Taking $ G_F^I(p)$ to be the $ G_F(p)$ in Eq. (55), we will have
$$\begin{array}{rcl}
G_r(p)&=& i /\{p^2-m^2+ i\varepsilon p^0+ {\rm Re}\tilde{K}(p)-\tilde{H}(p)+ 
                           i \eta(p^0)[\tilde{S}^I(p)+ {\rm Im}\tilde{K}(p) ]\}  
\\
&=& i /\{p^2-m^2+ i\varepsilon p^0+ {\rm Re}\tilde{K}(p)-\tilde{H}(p)+ 
                             i\sinh(\beta p^0/2)\tilde{R}(p)\} ,  \\
G_a(p)&=&-[G_r(p)]^*,
\end{array}\eqno(60)$$
which are respectively the expressions for retarded and advanced two-point
functions.
\section{Conclusions}
\indent By direct calculations, we have proven the equivalence of the whole
two-point function to one-loop order including the causal, retarded and
advanced one in the imaginary-time and the real-time formalism of thermal
$\lambda\varphi^3$ field theory and all the results are expressed in the
real-time formalism.  We find that the key-point of completing the proof is
carefully to consider and separate the imaginary part of the zero-temperature
loop integral.  The similar discussions can be generalized to the other theories
e.g. $\lambda \varphi^4$ field theory, QED and the calculations to higher loop
orders. In particular, it will be very useful for us to reexamine the
equivalence problem of the propagators for scalar bound states in a NJL model
in the two formalisms of thermal field theory.

\end{document}